\documentclass[aps,pra,reprint,groupedaddress]{revtex4-2}
\usepackage{amsmath}
\usepackage{graphicx}
\usepackage[dvipsnames]{xcolor}
\newcommand{\accVarphi}{A}
\newcommand{\change}[1]{{\color{magenta}#1}}
\renewcommand{\change}[1]{#1}
\usepackage[normalem]{ulem}

\makeatletter
\newcommand{\href@noop}[2]{#2}
\newcommand{\Eprint}[2]{\href{#1}{#2}}
\makeatother
\usepackage[colorlinks=true,linkcolor=DarkOrchid,citecolor=Plum,urlcolor=CadetBlue,hyperfootnotes=true]{hyperref}

\begin{document}

\title[Universal limit]{Limit on spatial quantum superpositions
with massive objects due to phonons}

\author{Carsten Henkel}
\email[]{henkel@uni-potsdam.de}
\affiliation{University of Potsdam, Institute of Physics and Astronomy, Germany}

\author{Ron Folman}
\email[]{folman@bgu.ac.il}
\affiliation{Department of Physics, Ben-Gurion University of the Negev, Beer Sheva, Israel}

\date{27 September 2024}


\begin{abstract}\noindent
It has been a long-standing goal to bring massive objects into a superposition of different locations in real space, not only to confirm quantum theory in new regimes, but also to explore the interface with gravity.
The main challenge is usually thought to arise
from forces or scattering due to environmental 
\change{
fields and particles%
} 
that decohere the large object's wave function into a statistical mixture.
We unveil a decoherence channel which cannot be eliminated by improved isolation from the environment.
It originates from sound waves within the object, which are excited as part of any splitting process and carry partial “\emph{Welcher Weg}” information. This puts stringent constraints on future spatial superpositions
of large objects.
\end{abstract}

\maketitle

\section{Introduction}

The superposition principle is one of the basic principles of quantum theory.
Spatial quantum superpositions have so far been tested only with small systems, from photons \cite{Aspden2016} and elementary particles \cite{Frabboni2012, Rosa2012, Rauch2015, Sala2019} to atoms \cite{Cronin2009, Rasel2019, Keil2021, Margalit2021} and molecules \cite{Fein2019, Shayeghi.2020}.
Quantum phenomena with massive objects such as entanglement \cite{Kotler.2021, MercierdeLepinay.2021} and superposition \cite{Schrinski.2023, Bild.2023} have been experimentally realized, but massive-object spatial superpositions remain a long-standing sought-after goal \cite{Pino2018, Wan2016, Millen2020, Kialka2022, Pedernales.2021, Neumeier.2024, Zhou.2024}.
Achieving a spatial superposition is important not only in order to verify quantum theory in new regimes \cite{Das.2024}, but also in order to probe the quantum-gravity interface \cite{Margalit2021, Bose2017, Marletto2017, Marshman2020,  Marletto2020, Carney2021, Streltsov.2022, Anastopoulos.2022}.
In addition, such an experiment may test exotic theories \cite{Diosi1989, Marshall2003, Romero-Isart.2011, Fuentes2018, Howl2019}, and may even enable new technology \cite{Rademacher2020}.

To create 
a superposition state, one needs to start with a single object and then apply a splitting force so that 
it splits in real space into two exact copies, wave packet, mirroring the original object, but with opposite momentum components.
Quantum theory permits us to calculate the properties of these individual wave packets as if they were real objects.
This is crucial 
to evaluate the contrast of an interference pattern that may be formed if the two wave packets are eventually joined (thus closing a loop in space-time).
Such an interference pattern is the only way to prove that indeed a coherent quantum superposition was formed.
To give an example of the fundamental implications, we recall that because Stern and Gerlach (SG) did not recombine the two beams in their seminal experiment \cite{Gerlach1922} to show an interference pattern, a century of debate ensued of whether SG splitting can indeed create spatial quantum superpositions (see \cite{Margalit2021} and references therein).
This debate has only recently been experimentally put to rest \cite{Amit.2019,Margalit2021,Keil2021}.

In this paper, we calculate the orthogonality that develops, as a massive object is brought into a spatial superposition.
When orthogonality becomes strong, there is no interference pattern, and an operational proof for a spatial superposition is impossible.
To mitigate orthogonality, two issues have to be addressed:
First, the imperfect closing of the interferometer loop (imprecise recombination) which has been coined the “Humpty-Dumpty effect” \cite{Englert1988,Schwinger1988,Scully1989}.
For minimal orthogonality, a complete overlap (identity) of position, momentum, rotation \cite{Japha.2022}
and even wave packet size and shape, is required at the recombination point.
The second challenge is environmental decoherence \cite{Joos.1985,Zurek1991,Chang2009,Romero-Isart.2011,Albrecht2014,Bateman2014,Kamp2020,Schut2022,Kialka2022}, whereby the object couples to an environment whose many degrees of freedom (DoF) monitor the wave packets and entangle with them, eventually performing a projective measurement.
However, advances in the technology of isolation could in future suppress such decoherence.

Here, we assume the above two channels do not produce orthogonality, and  focus our attention on 
DoF which are unique to massive objects, phonons.
This decoherence channel is internal to the object, and consequently improved isolation would not help.
Phonons are sound waves
where atoms in the bulk move relative to each other.
{In the context of Mössbauer spectroscopy, their role has been appreciated
as providing a broad background spectrum, as opposed to the recoil-less
(or “zero-phonon”) line. 
The model we develop is inspired by the formalism pioneered by 
W. E. Lamb where the force acting 
\change{
on%
}
a single impurity
nucleus is driving the atomic constituents of an object, as expanded
over the set of phonon normal coordinates \cite{Lamb.1939,Lipkin.1960b}.}
When the above noted splitting force acts on the atoms in the object,
the crucial question is, how homogeneously?
If different atoms feel a slightly different force, phonons will be excited.
Since, in order to achieve a splitting, the forces acting on the two wave packets 
must be of opposite sign (direction), the
generated
phonons 
have a different sign or phase, and this gives rise to orthogonality
\cite{Lipkin.1961}.
We estimate in particular the orthogonality due to a large number of phonon 
modes. 
%
%
While the general idea of this “internal environment” has been noted
before \cite{Leggett.2002}, our model provides a detailed 
\change{
quantitative%
}
prediction
for the phonon impact on orthogonality in the context of splitting and
recombining massive objects.

Last, we note, that while other hypothesized internal decoherence mechanisms require extensions to standard quantum theory (e.g., the Diosi-Penrose gravitationally induced collapse model or
continuous spontaneous localization
models \cite{Schlosshauer.Book, Gasbarri2021}), the mechanism presented here does not require any such extension.

\section{Model}

\subsection{Excitation of phonons}

As a general model for the process of splitting a nano-particle, 
independent of a specific material or interaction,
we consider a potential $\alpha_i U(x)$ with some field intensity $U$ 
and a coupling $\alpha_i$ to a specific atom in the lattice.
The force acting on the atom with index $i$ is thus $F_i=-\alpha_i U'(x)$, where the derivative is evaluated at the position of the atom. 
The equation of motion for the atoms can be written
\begin{equation}
m_i \ddot x_i + \sum_j K_{ij} x_j = F_i
\,.
\label{eq:Newton-1}
\end{equation}
Here, the atom labelled $i$ is displaced by $x_i$ from its equilibrium position ($x_{0i}$), and the matrix elements $K_{ij}$ specify the interatomic spring constants.
To avoid phonons from being excited we require that the force is homogeneous.
This means that $\delta U'$, describing how far from constant is the potential gradient across the extent of the object, and $\delta\alpha$, the variation in
the coupling to the potential gradient, should be very small. Specifically, $\delta\alpha$ is a measure of to what degree the atoms in the material are all the same
(chemical and isotopic purity),
and to what degree their density and orientation are the same across the object (e.g., no geometrical defects).

While the force variation $\delta F$ should be represented by a power spectrum, 
we are only interested in its intensity
on spatial scales (inverse $k$ vectors) from the object size to the atomic interspacing.
An example from the atomic scale
may be an inhomogeneous material whereby spin contaminations producing a Stern-Gerlach force are randomly scattered throughout the object (e.g., 
nitrogen-vacancy centers within the spinless $^{12}{\rm C}$ diamond lattice \cite{Margalit2021}).
This is the model we are going to work with in the following,
based on the force correlation function
\begin{equation}
\overline{ \delta F_i(t) \delta F_j(t') } = 
\delta_{ij} m_i m_1 
\frac{ \delta F^2 }{ F^2 } 
a(t) a(t')
\,.
\label{eq:atomic-white-noise}
\end{equation}
Here, $\delta F / F$ is the relative amplitude of the force inhomogeneities,
and $a(t)$ is the acceleration per ``average atom'' with reference mass $m_1$.

Any finite-range spatial correlation in the force $F_i$ would lower its power spectrum at short scales, reducing the excitation of phonons with the corresponding wavelengths.
On a scale comparable to the object size, for example, we may consider
an inhomogeneous force causing a collective stretching of the object. 
Such an effect would originate from an inverse harmonic potential \cite{Romero-Isart.2017} or from tidal forces due to gravity. 
Let us briefly note that in principle, since for heavy objects one would need to apply a force for long durations $\Delta t$ to get a meaningful spatial splitting, the Fourier transform of such a long pulse results in low frequencies. 
Consequently, extreme $k$ vectors comparable to the inverse of the atomic interspacing, will be relatively suppressed as they require high frequencies. Specifically, if we take the velocity of sound in the bulk to be of the order of $c=1000\,{\rm m/s}$ and the inter-atomic spacing to be 
{$1\,{\rm \AA}$}, the required frequencies are 
{$10\,{\rm THz}$}. 
This ``Debye frequency'' would require pulse durations of the applied force shorter than pico-second and concomitant large accelerations, unrealistic for large-object splitting.
For the same reason of not considering ultra-short $\Delta t$, we will be neglecting the optical phonon branch. 
Hence, in the following we will be calculating a lower bound on the produced orthogonality. 

\begin{figure*}[htbp]
   \centering
   \includegraphics*[width=0.7\textwidth]{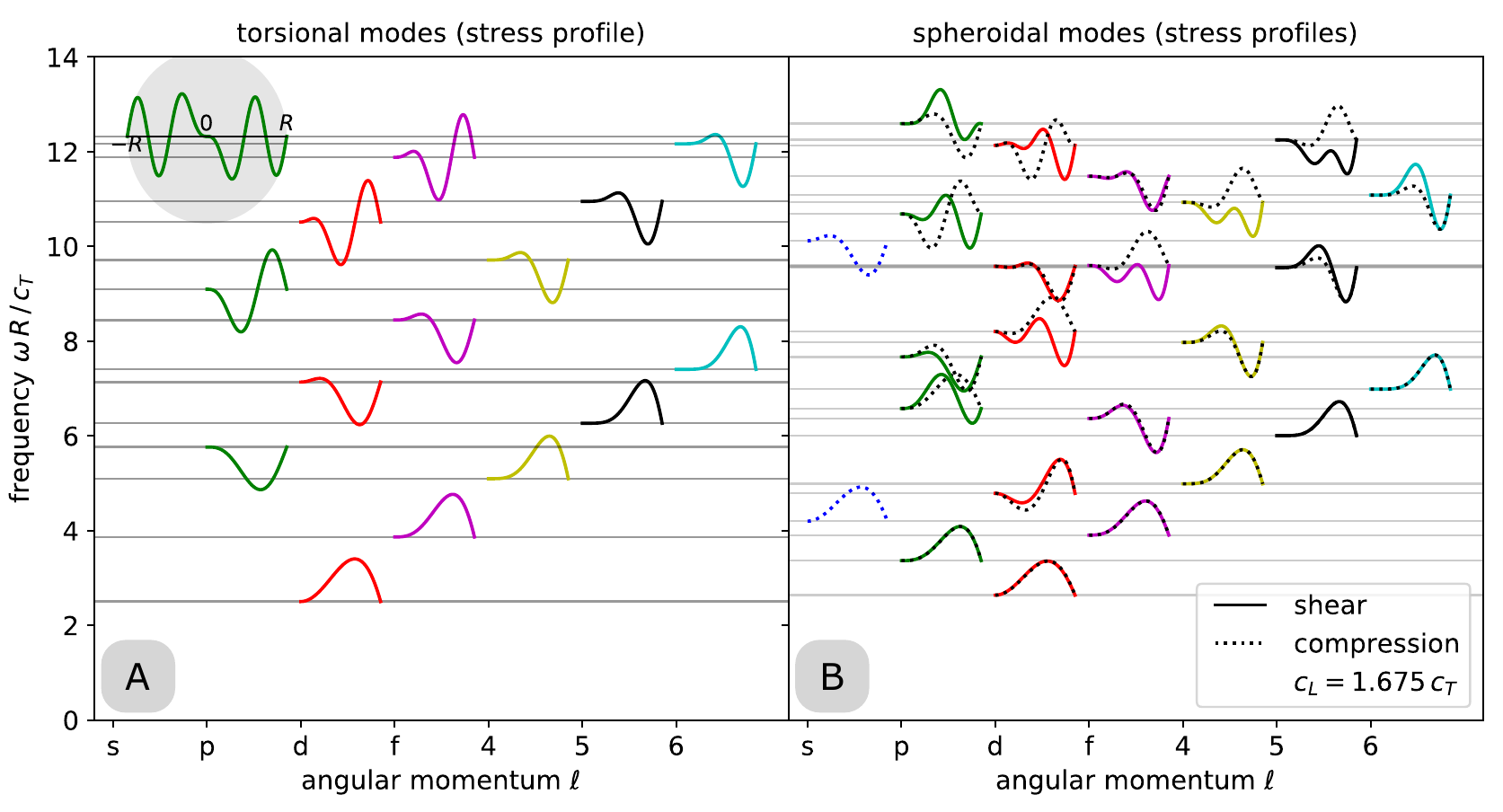}
   \vspace*{-2ex}
   \caption[]{%
Mode functions for sound waves in a solid sphere, according to \cite{Lamb.1882} and \cite{GonzalezBallestero.2020b}.
The speeds of shear and compressional waves are representative for silicon
\cite{Wang.2004,Norris.2006}.
For a sphere with diameter $2R = 1\,\mu{\rm m}$, the torsional d-waves (A) provide the fundamental tone at $\omega/2\pi \approx 2.501\, c_T/(2\pi R) \approx 4.26\,{\rm GHz}$,
while spheroidal d- and p-waves (B) are slightly higher in frequency.
The plotted scalar functions provide the radial profile of the compressional and shear stress from $0$ to $R$. They vanish on the sphere surface which is stress-free.
In the top left corner, one mode is shown across the full object diameter. The spectrum is rather irregular making
coherent
phonon control (e.g., in order to reverse their evolution) 
challenging 
for large objects.
}
   \label{fig:sphere-modes}
\end{figure*}

In Fig.\:\ref{fig:sphere-modes} we present the phonon modes we are taking into account.
For a solid sphere, they have been computed by Horace Lamb,
with frequencies related to zeros of the spherical Bessel functions \cite{Lamb.1882, GonzalezBallestero.2020b}.
They depend only on the ratio $c / D$, namely the speed of sound over the diameter.
A simple re-scaling thus gives the spectrum for a sphere of any size.
\change{
Phonon spectra for objects with a broad range of shapes are available 
on the web site of L. Saviot \cite{SaviotWeb}.%
}

While Fig.\:\ref{fig:sphere-modes} arises from a continuum description, 
we can also introduce normal (phonon) modes for the lattice model of Eq.\,(\ref{eq:Newton-1}) \cite{Lamb.1939, AshcroftMermin.Book}.
They are found as eigenvectors $u^{k}_i = u^{k}(x_{0i})$ of the dynamical matrix:
\begin{equation}
\sum_j \frac{ K_{ij} }{ \sqrt{ m_i m_j } } u_{j}^k = \omega_k^2 \,
\change{
u_{i}^k%
}
\,,
\label{eq:dynamical-matrix}
\end{equation}
where $\omega_k$ is the phonon frequency.
\change{
Here, the index $k$ labels the eigenvalues $\omega_k$, and the eigenvector ${\bf u}^{k}$ has as components the dimensionless displacement $u^{k}(x_{0i})$ of atom number $i$ in the object relative to its equilibrium position $x_{0i}$.%
}
All {eigenvectors} can be chosen real and orthogonal, and be normalized according to $\sum_i u^k_{i} u^l_{i} = \delta_{kl}$.
They provide an expansion of the displacements of Eq.\,(\ref{eq:Newton-1}) 
into normal modes, 
\begin{equation}
x_i(t) = \sum_k q_k(t) u^k_i / \sqrt{ m_i }
\label{eq:def-mode-expansion}
\end{equation}
with mass-weighted mode amplitudes $q_k(t)$ (unit: length times root of mass).
The center-of-mass {(CoM)} mode corresponds to a zero-frequency eigenvector $u^{0}_{i} = \sqrt{m_i/M}$ with the total mass $M$, 
\change{
where according to Eq.\,(\ref{eq:def-mode-expansion}), all atoms are subject to the same displacement.%
}
This mode exists because of global translation invariance of the potential energy, i.e., $\sum_j K_{ij} = 0$.
\change{
Using the expansion of Eq.\,(\ref{eq:def-mode-expansion}),%
}
the energy of the lattice takes the diagonal form
\begin{equation}
H = \sum_i \frac{m_i}{2} \dot x_i^2 + \sum_{ij} \frac{K_{ij}}{2} x_i x_j
= \frac{1}{2}\sum_k \big( \dot q_k^2 + \omega_k^2 q_k^2 \big)
\,.
\label{eq:energy-per-mode}
\end{equation}
Project the equation of motion~(\ref{eq:Newton-1}) onto the normal mode $k$ to find
\begin{align}
\ddot q_k + \omega_k^2 q_k = f_k(t)
&:= \sum_i \frac{ u^k_i F_i }{ \sqrt{m_i} }
\,,
\label{eq:def-fk}
\end{align}
where $f_k(t)$ is the force acting on the mode.
This is the equation of a driven oscillator with the solution (depicted in Fig.\:\ref{fig:orthogonality})
\begin{equation}
\begin{aligned}
q_k(t) = \ & q_k(0) \cos\omega_k t + \dot q_k(0)\frac{\sin\omega_k t}{\omega_k}
\\
& + \int_0^{t}\!{\rm d}t'\frac{ \sin\omega_k (t - t')}{\omega_k} f_k(t')
\,.
\end{aligned}
\label{eq:solution_qk}
\end{equation}
%
%
In the phase space spanned by the canonical coordinates $q_k$ and $\dot q_k$
this combines a free rotation by the angle $\omega_k t$ 
and a translation.
The mode amplitude $q_k$ is shifted by the integral term in Eq.\,(\ref{eq:solution_qk}), while the in-quadrature component $\dot q_k$ is displaced by
\begin{equation}
\int_0^{t}\!{\rm d}t'\cos[\omega_k (t - t')] f_k(t')
=
\tfrac{1}{2}\Delta \dot q_k(t)
\,.
\label{eq:displacement-in-momentum}
\end{equation}
We collect these shifts into $\Delta q_k(t), \Delta \dot q_k(t)$.
Imagine that along the two arms of the interferometer, forces with opposite
signs are applied: $\Delta q_k(t)$ is then the (mass-weighted) distance
between the two wave packets with respect to the phonon coordinate.

\begin{figure}[bhtp]
   \centerline{%
   \includegraphics*[width=1.05\columnwidth]{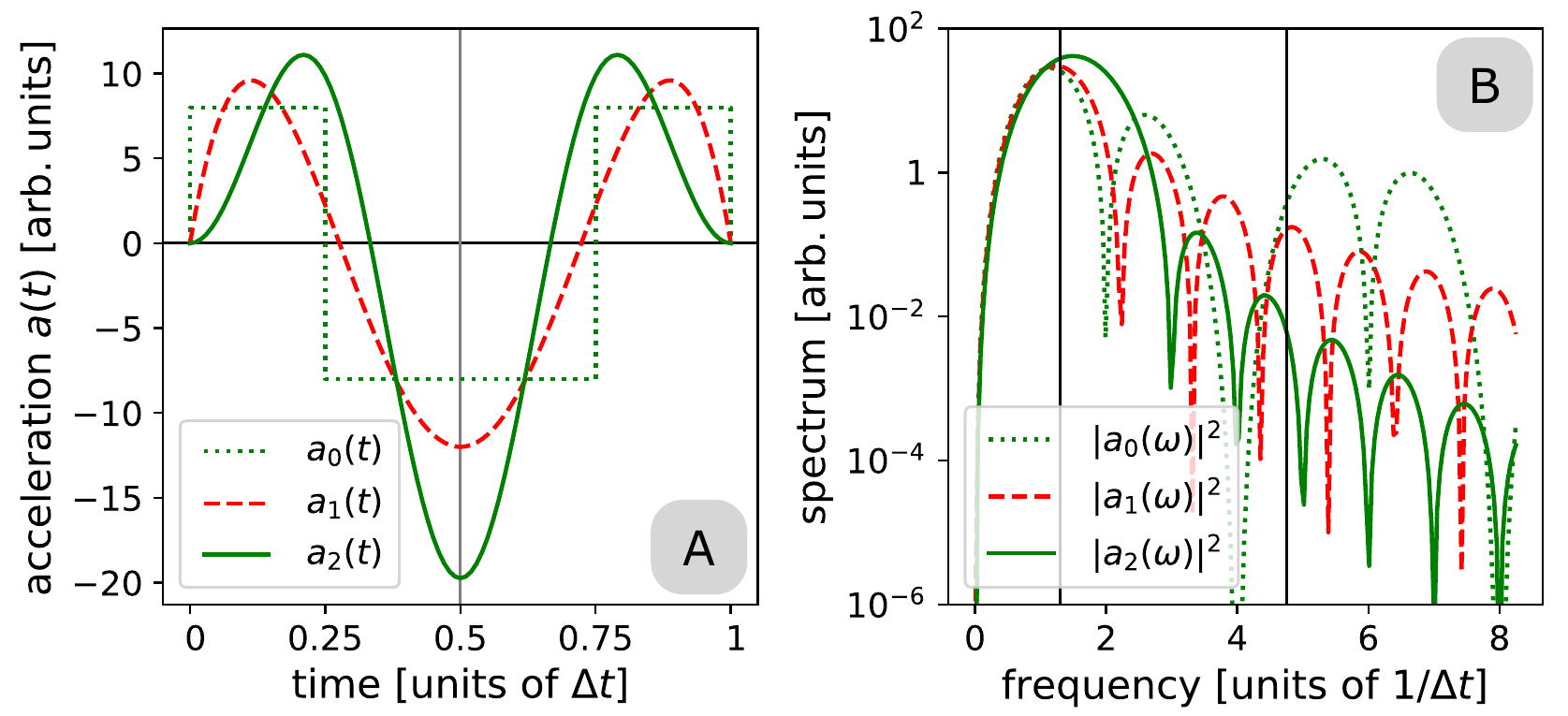}%
   }%
   \vspace*{-2ex}
   \caption[]{%
Profiles for the splitting force, represented by $a_{n}(t)$, $n = 0, 1, 2$.
They have in common the same maximum spatial splitting but differ in the degree of adiabaticity.
The 
{profile $a_0(t)$ is rectangular} (the force flips sign instantaneously at $t = (\frac{1}{4}, \frac{3}{4})\Delta t$), so less adiabatic, while the other two are more adiabatic.
(A) 
The force first splits, then reverses and stops again
the relative motion for a full overlap in position and momentum.
(B) A Fourier transform of the force profiles in time, where as expected, the rectangular profile gives much more weight to higher (nonadiabatic) frequencies.
The vertical lines mark the phonon frequencies considered in Fig.\:\ref{fig:orthogonality}.
}
   \label{fig:force-profiles}
\end{figure}

Representative examples are given in Figs.~\ref{fig:force-profiles},~\ref{fig:orthogonality}.
We assume a general closed-loop symmetric configuration in which four equal (up to a sign) forces are applied, each of duration $\Delta t/4$, for splitting, stopping the relative motion, accelerating back, and again stopping the relative motion for a full overlap in position and momentum [see Fig.\:\ref{fig:force-profiles}(A)]. As noted above, the interaction giving rise to the force may be of any type, such as light based \cite{Cronin2009, Neumeier.2024} or magnetic based (as in our previous work \cite{Margalit2021, Henkel2022}).

\begin{figure}[htbp]
   \centerline{%
   \includegraphics*[width=1.1\columnwidth]{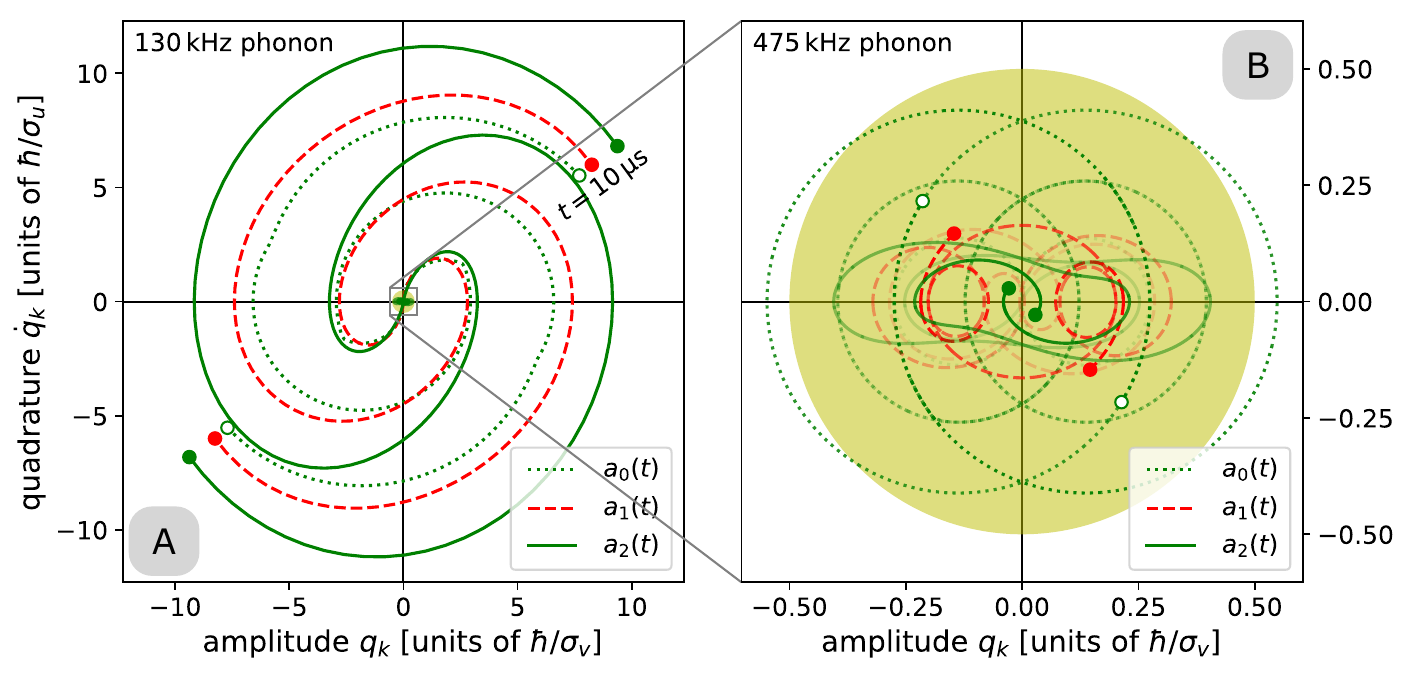}%
   }
   \caption[]{%
Illustration of phonon orthogonality: the amplitude of a given phonon mode evolves under the applied splitting force pulse, proportional to the projection of the inhomogeneous force profile onto the spatial mode pattern [Eq.\,(\ref{eq:def-fk})].
The plots depict the trajectories in phase space for forces with opposite signs: on the horizontal axis the mode's excitation amplitude $q_k$ and on the vertical the corresponding {in-quadrature component} $\dot q_k$.
The parameters are taken for a pulse duration of $10\,\mu{\rm s}$, a targeted maximum spatial splitting of $0.1\,\mu{\rm m}$ (maximum acceleration of the order of $10^3\,{\rm m/s}^2$), $\delta F/F = 1\%$, and a temperature of $T = 4\,{\rm K}$ (de-Broglie wavelength of a single atom $\approx 0.1\,{\rm nm}$).
The quadratures $q_k, \dot q_k$ are made dimensionless by dividing by the corresponding coherence lengths in the initial thermal state [Eq.\,(\ref{eq:th-eq-sigma-uv})].
The yellow circle of unit diameter corresponds to the coherence area of the initial state: outside it, the contrast drops below $1/e$.
(A) Relatively low frequency: at the end of the applied force pulse, the phonon excitation is so large (dots) that the two interferometer arms can be distinguished on the scale of the coherence of the initial state (yellow circle). A single low-frequency mode in the object's phonon spectrum would ruin the interference contrast.
(B) Higher frequency: during the applied pulse, the mode follows a more complex dynamics (the initial stage is faded out for better visualization), but the final excitation is much weaker.
Still, exciting many high-frequency modes also reduces the contrast significantly.
The line styles correspond to the acceleration protocols of Fig.\:\ref{fig:force-profiles} with increasing degree of adiabaticity.
The chosen frequencies $\omega_k \, \Delta t / 2\pi = 1.3, 4.75$ {are marked by} the vertical lines in Fig.\:\ref{fig:force-profiles}\,(B).
While the more adiabatic protocols give a smaller excitation at the higher frequency, 
the lower frequency is chosen such that, incidentally, the reverse applies.
}
   \label{fig:orthogonality}
\end{figure}

Since Eqs.\,(\ref{eq:solution_qk}, \ref{eq:displacement-in-momentum}) 
are a Fourier transform of the applied pulsed force, 
we note that slow pulses (large $\Delta t$) help to minimize phonon excitation.
This is, however, not simple to implement, as it is well known that the larger the mass of an object, the harder it is to isolate it from the environment and the larger the decoherence rate. 
But we assume that future technology will indeed enable near-perfect isolation.
The temporal acceleration profiles $a(t)$ and their spectra 
in Fig.\:\ref{fig:force-profiles}(B) illustrate the
impact of adiabaticity: a slow and smooth rise of the splitting force reduces the excitation of high-frequency phonons. 
The dips in the spectra may suggest 
a scheme where the applied pulses are taylored to avoid the excitation of low-lying
modes. 
This is, however, not quite realistic, as a glance at the irregular spectrum of Fig.\:\ref{fig:sphere-modes} shows. 
The details of the phonon spectrum will also depend, of course, on nanoscopic features of the object (aspect ratio, protrusions, roughness) that are difficult to control in fabrication and challenging to characterize for a given object. 
In the following, we thus adopt a statistical approach where the unknown 
forces on the atomistic scale are described by the correlation 
function in Eq.\,(\ref{eq:atomic-white-noise}).

\subsection{Orthogonality of phonon quantum states}
\label{s:quantum-orthogonality}

The amplitude of each phonon mode is, in our model, proportional to the sign of the applied force: it thus carries “\emph{Welcher Weg}” information.
We quantify in this section by how much the interference contrast is reduced due to this information being potentially available.
An example for the formation of orthogonality is presented in Fig.\:\ref{fig:orthogonality}
which shows a phase-space diagram of the trajectories for phonon mode amplitudes in the two wave packets.
As can be seen, once separated, they never fully overlap again on the scale of their initial coherence area (yellow disks).
This happens although the loop is closed in the standard phase space of CoM position and momentum, as can be seen from Fig.\:\ref{fig:force-profiles}.
We emphasize that the orthogonality discussed in this work arises from unclosed loops in the unique phase space of phonon amplitudes. 
This is more subtle than the excitation of phonons with opposite $k$-vectors (propagating to the left or right, according to the arm of the interferometer).
Indeed, orthogonality arises due to the complex amplitudes of any phonon mode, be it a single one with a definite sign of its ``momentum'' (not to be confused with the momentum operator $\hat{\dot{q}}_k$ canonically conjugate to $\hat{q}_k$).

In quantum theory, the transformation in Eq.\,(\ref{eq:solution_qk}) of the operator-valued phonon variables $\hat q_k$ and $\hat{\dot q}_k$
combines a rotation (free evolution at $\omega_k$) 
with a displacement.
The rotation is generated by the quantized Hamiltonian in Eq.\,(\ref{eq:energy-per-mode}), 
while the displacement operator has the form $\hat{D}_k$ \cite{VogelWelsch.Book}
\begin{equation}
\hat{D}_k(t) = \exp\frac{i}{2\hbar}\big[
  \Delta \dot q_k(t) \, \hat q_k
- \Delta q_k(t) \, \hat{\dot q}_k \big]
\,.
\label{eq:displacement-operator}
\end{equation}
Here the hats mark the canonically conjugate operators, 
and the $\Delta q$'s are parameters. 
Consider now a setting in the Schrödinger picture 
that starts from a stationary state $\vert \psi_k(0)\rangle$ 
for the normal mode $k$. 
It is invariant under rotation, and the applied force generates the state
\begin{equation}
\vert \psi_k(t)\rangle = \hat{D}_k(t) \, \vert \psi_k(0)\rangle
\,,
\label{eq:}
\end{equation}
up to a phase factor independent of $\hat{D}_k(t)$.
{The interferometer scheme prepares} a spatially separated wave packet if, for example, the sign of the force is controlled by a quantum variable like a spin. In Refs.\,\cite{Amit.2019,Margalit2021}, experiments were done with atoms carrying an unpaired electron and a magnetic gradient force proportional to some projection of the spin magnetic moment (Stern-Gerlach effect) \cite{Gerlach1922,Keil2021,Marshman2022}.
Depending on the spin projection (with signs $\pm$), we thus get displaced states
\begin{equation}
\vert\psi^\pm(t)\rangle = \bigotimes_k \hat{D}^{(\pm)}_k(t) \vert\psi(0)\rangle
\,,
\label{eq:displaced-state}
\end{equation}
where the superscript flips the sign of the displacement 
\change{
parameters%
}
in Eq.\,(\ref{eq:displacement-operator}).
The (tensor) product is taken over all phonon modes of the particle, assuming that the initial state is factorized. 
This is a good approximation when phonon anharmonicity is weak.

At the end of the splitting scheme, after spatial recombination and just before detection, the object's quantum state is the multi-mode superposition
\begin{equation}
\vert\psi(\Delta t)\rangle = \frac{1}{\sqrt{2}}\big(
\vert\psi^+(\Delta t)\rangle + {\rm e}^{ {\rm i}\phi } \vert\psi^-(\Delta t)\rangle
\big)
\,.
\label{eq:superposition}
\end{equation}
This state produces interference fringes as the phase difference $\phi$ is scanned. Their normalized contrast turns out to be the mode product \cite{Lipkin.1960b}
\begin{align}
C & = \vert \langle\psi^-(\Delta t) \vert \psi^+(\Delta t)\rangle \vert
\nonumber
\\
& = \prod_k \Big\vert \mathop{\rm Tr}\big[
\hat{D}^{(+)}_k(\Delta t) \vert\psi_k(0)\rangle \langle \psi_k(0) \vert
\hat{D}^{(-)\dag}_k(\Delta t)
\big] \Big\vert
\label{eq:Bloch-formula-0}
\\
&= \prod_k \big\vert \mathop{\rm Tr}\big[
\hat{D}^{(+)2}_k(\Delta t) \rho_k(0)
\big] \big\vert
\,.
\nonumber
\end{align}
This formula illustrates that for any single mode, the non-closing of the loop reduces the interference contrast.
Note that the third line of Eq.\,(\ref{eq:Bloch-formula-0}) generalizes the contrast to an initial state $\rho_k(0)$ that is mixed, e.g.\ the thermal equilibrium state for each phonon mode. 
\change{
For a harmonic lattice, its equilibrium state is a (separable) tensor product and compatible with the normal-mode product in Eq.\,(\ref{eq:Bloch-formula-0}).%
}

To evaluate the traces in Eq.\,(\ref{eq:Bloch-formula-0}), it is expedient
to use the Wigner representation $W_k(q, \dot q)$ of the density operator
\cite{Schleich.Book}.
It leads to the so-called {Bloch} formula \cite{Milonni.Book, Bloch.1932}
for a harmonic mode. 
If its equilibrium Wigner function is the double Gaussian
\begin{equation}
W_k(q, \dot q) = N \, \exp\Big( {-} \frac{q^2}{2\sigma_u^2} - \frac{\dot q^2}{2\sigma_v^2} \Big)
\,,
\label{eq:Wigner}
\end{equation}
the trace becomes the double Fourier transform of $W_k$
\begin{equation}
\mathop{\rm Tr}\big[
\hat{D}^2_k(\Delta t) \rho_k(0)
\big]
= \exp\Big[
{-} \frac{ 1 }{ 2\hbar^2 }
\big( 
\Delta \dot q_k^2 \sigma_u^2 
+ 
\Delta q_k^2 \sigma_v^2 
\big)
\Big]
\,,
\label{eq:contrast-exp-q2}
\end{equation}
This formula is {analogous to the 
Lamb-Mössbauer factor for the ratio of elastic to inelastic
neutron scattering, similar to the Debye-Waller factor
\cite{Lamb.1939, Lipkin.1960b, Lipkin.1961}.}

The thermal equilibrium Wigner function in Eq.\,(\ref{eq:Wigner}) contains variances given by \cite{VogelWelsch.Book, Schleich.Book}
\begin{equation}
\sigma_u^2 = \frac{ \hbar }{ 2 \omega_k } \coth\tfrac12 \beta \omega_k
\,,\qquad
\sigma_v^2 = \frac{ \hbar \omega_k}{ 2 } \coth\tfrac12 \beta \omega_k
\,,
\label{eq:th-eq-sigma-uv}
\end{equation}
where $\beta = \hbar/k_BT$ involves the internal temperature $T$ 
of the $k$'th normal mode.
Combining Eqs.\,(\ref{eq:Bloch-formula-0}, \ref{eq:contrast-exp-q2},
\ref{eq:th-eq-sigma-uv}), we get the logarithm of the interference contrast 
as a sum over phonon modes
\begin{equation}
\log C = - \sum_k
\frac{ \coth\tfrac12 \beta \omega_k }{\hbar\omega_k}
\bigg\vert
\int\limits_0^{\Delta t}\!{\rm d}t\,
{\rm e}^{ {\rm i} \omega_k t }
f_k(t)
\bigg\vert^2
\,,
\label{eq:log-contrast-mode-sum}
\end{equation}
where $f_k(t)$ was defined by Eq.\,(\ref{eq:def-fk}).

As a simple check, consider the case that the force acting on the object's $i$th atom is gravity, $F_i = m_i g$ with a spatially homogeneous acceleration $g$.
This is admittedly not very practical for a controlled splitting, but it is interesting to note that the {projected force on mode $k$ is}
\begin{equation}
f_k = \sum_i g \, u^k_i \sqrt{m_i} = g \, \sqrt{M} \, \delta_{k,0}
\,,
\label{eq:}
\end{equation}
where $M$ is the object's total mass.
Such a force couples \emph{only} to the CoM\ mode $k = 0$, a nice illustration of the equivalence principle
{and sometimes called the “diver's theorem” \cite{Leggett.2002}.}
The CoM term in the contrast of Eq.\,(\ref{eq:log-contrast-mode-sum}) can be made finite by assuming that the CoM\ coordinate is initially distributed with a certain width $\sigma_x$.
This fixes the variances in Eq.\,(\ref{eq:th-eq-sigma-uv}) 
to $\sigma^2_u = \sigma_x^2 M$,
and to $\sigma_v^2 = k_B T$ with the kinetic (CoM) temperature $T$.
The contrast reduction can be expressed in terms of the \emph{final differential displacements} $\Delta x_f$, $\Delta p_f$ in CoM\ position and momentum
\begin{equation}
\log C_{\rm CoM}
= - \frac{ \Delta x_f^2 }{ 2 \lambda_T^2 }
  - \frac{1}{2} \left(
\frac{ \Delta p_f\, \sigma_x }{ \hbar } \right)^2
\,,
\label{eq:cm-log-contrast}
\end{equation}
with the thermal de-Broglie wavelength $\lambda_T = \hbar / (M k_B T)^{1/2}$ of the \emph{entire} object.
We recall that $\tfrac{1}{2}\Delta x_f$ is the displacement in one arm of the interferometer from its initial position; for a wave packet split with symmetrically opposite forces, $\Delta x_f$ measures its spatial non-perfect overlap at the final time $t = \Delta t$.
Equation~(\ref{eq:cm-log-contrast}) illustrates the accuracy that is required in ``closing the loop'' in phase space \cite{Schwinger1988}.
(In practice, the overlap has also to be optimized with respect to the width of the wave packets and the rotation angles of the object \cite{Japha.2022}.)

Returning to our model for phonon-related orthogonality, 
the final step is to estimate the
contrast in Eq.\,(\ref{eq:log-contrast-mode-sum}) by taking its average over the
force correlation function. 
This gives, for example,
\begin{align}
\overline{ f_k(t) f_k(t') } 
&= 
\sum_{ij}
  u^k_i u^k_j
\left(
  \sqrt{m_i m_j} + \delta_{ij} m_1 \frac{ \delta F^2 }{ F^2 } \right) 
  a(t) a(t')
\nonumber\\
&=
M \delta_{k,0} \, a(t) a(t')
+ m_1 \frac{ \delta F^2 }{ F^2 } a(t) a(t')
\,,
\label{eq:average-force-mode-k}
\end{align}
where the normalisation of the phonon modes [see after Eq.(\ref{eq:dynamical-matrix})] has been used. 
The first term in Eq.\,(\ref{eq:average-force-mode-k}) describes the excitation of the CoM: as mentioned above, we are assuming that this yields a perfect overlap.
The second term with $\delta F/F$ gives the average excitation of any other phonon mode.
Introducing the Fourier transform $\tilde{a}( \omega )$ of the
acceleration $a(t)$, taken over the finite experiment 
time $t = 0 \ldots \Delta t$,
the average contrast finally takes the form
\begin{equation}
\log C \lesssim -
\frac{ \delta F^2 }{ F^2 } m_1
\sum_{k \ne 0} \frac{ \coth\tfrac12 \beta \omega_k }{ \hbar \omega_k }
\vert \tilde{a}( \omega_k ) \vert^2
\,.
\label{eq:log-contrast-2}
\end{equation}
\change{
Here, the inequality takes into account the orthogonality 
due to any phonon modes not included in the model (e.g., the optical branches).%
}
This is the main result of the paper.
Note that at this point, microscopic details like spatial mode patterns and polarisation drop out, and only the phonon spectrum $\{ \omega_k \}$ remains relevant.
\change{
It should be emphasized that our prediction in Eq.\,(\ref{eq:log-contrast-2}) for the contrast is more general than the Debye model, since no further assumptions about the phonon density of states are needed.%
}

\subsection{Summing over phonons}

In the following, we distinguish between two limiting cases
to evaluate the phonon sum in Eq.\,(\ref{eq:log-contrast-2}).
Only a few phonon modes are relevant if during the time $\Delta t$, sound (with speed $c$) has made many round trips across the size $L$, i.e., $c\, \Delta t \gg L$. 
This applies to ``small'' objects.
In the opposite case of a ``large'' object, many modes contribute. 
In that case, sound waves originating from the object's surface, for example, have not yet reached the other end when the wave packets recombine.

The two cases can also be rationalized by comparing the bandwidth $1/\Delta t$ of the force pulse to the ``fundamental tone'' $\omega_1 \sim \pi c / L$ of the phonon spectrum. 
The latter is a wave that fits within the boundaries of the object. 
It falls into the GHz range for typical condensed-matter values of $c$ and objects smaller than a micron, see Fig.\:\ref{fig:sphere-modes}.
Such objects are indeed ``small'' when the force pulse is longer than a nanosecond. (Typical numbers are in the range of $\Delta t \sim 10 \ldots 100\,\mu{\rm s}$, depending on the available accelerations and spin coherence times.)

The upper limit of the acoustic mode spectrum is of the order of the Debye frequency (at the Brillouin zone boundary of the phonon band structure), typically a few THz.
The temperature dependence embodied in the parameter $\beta$ in Eq.\,(\ref{eq:log-contrast-2}) is such that $\beta \approx 1.9\,{\rm ps}$ ($25\,{\rm fs}$) at $T = 4\,{\rm K}$ ($300\,{\rm K}$), respectively.
In most cases, the classical approximation $\coth\tfrac12 \beta \omega_k \approx 2 k_B T / (\hbar\omega_k)$ is therefore applicable.

\subsubsection{Few modes}
\label{s:few-modes}

The few-mode regime corresponds to the parameters above the slanted gray band in Fig.\:\ref{fig:results} below.
The fundamental tone $\omega_1$ of the object is then in the high-frequency tail of the Fourier spectrum of the applied pulse. 
The sum in Eq.\,(\ref{eq:log-contrast-2}) over eigenfrequencies converges rapidly with only a few terms contributing significantly.
In the phonon spectrum of a sphere, it is therefore sufficient to keep only the two lowest acoustic frequencies [torsional and spheroidal d-waves, see Fig.\:\ref{fig:sphere-modes}(B)] with their degeneracy (10 modes).
The power laws that emerge in the Fourier transform $\tilde{a}( \omega )$ at high frequencies [see Fig.\:\ref{fig:force-profiles}(B)] are then essential for the loss of contrast. 
They can be computed explicitly and are given in Appendix~\ref{a:a-omega-and-other-a's}, Eq.\,(\ref{eq:a0-spectrum}--\ref{eq:spectrum-acceleration}).

It is convenient to exhibit in the acceleration profile the maximum spatial splitting $\Delta x$ between the wave packets, according to
\begin{equation}
\tilde{a}( \omega ) =
\frac{ \Delta x }{ \Delta t } \accVarphi(\omega \Delta t)
\,,
\label{eq:scaling-a-omega}
\end{equation}
where $\accVarphi(\omega \Delta t)$ is a dimensionless function with a maximum of order unity.
With this, Eq.\,(\ref{eq:log-contrast-2}) becomes
\begin{equation}
\log C \lesssim -
10 \frac{\delta F^2}{F^2}
\frac{ \Delta x^2 }{ \lambda_1^2 }
\frac{ \vert \accVarphi( \omega_1 \Delta t ) \vert^2
    }{ (\omega_1 \Delta t)^2 }
\,.
\label{eq:log-contrast-3}
\end{equation}
Here, $\lambda_1$ is the thermal de-Broglie wavelength for a \emph{single} representative atom (mass $m_1$, kinetic temperature $T$).
{It has been} checked that Eq.\,(\ref{eq:log-contrast-3}) is in good agreement with a summation over the lowest approx.\ 200 eigenfrequencies.

\subsubsection{Many modes}
\label{s:many-modes}

In this limiting case,
%
the details of the phonon spectrum become irrelevant.
Indeed, the sum over modes in Eq.\,(\ref{eq:log-contrast-2})
may then be evaluated by integrating.
Keeping three acoustic branches with an average speed of sound $\bar c$ and adopting again the classical approximation, we find 
\begin{equation}
\log C \lesssim
- \frac{ \delta F^2 }{ F^2 }
\frac{ 3 k_B T m_1 V }{ \hbar^2 \pi \, \bar c^3 }
\int_0^\infty\!\frac{{\rm d}\omega}{\pi} \, \vert \tilde{a}( 
\change{
\omega%
} 
) \vert^2
\,.
\label{eq:large-object-limit}
\end{equation}
Note that this expression only depends on the object volume $V$, independent of its shape, and of course, as noted, it is not dependent on any specific potential.
\change{
We shifted the upper frequency limit of the integral (which is physically of the
order of the Debye frequency) to infinity, assuming that the applied acceleration
sequence is sufficiently long. 
(See the estimate above, just before Sec.\:\ref{s:few-modes}.)%
}
The last integral can be transformed into the time domain
\begin{equation}
\int_0^\infty\!\frac{{\rm d}\omega}{\pi} \, \vert \tilde{a}( 
\change{
\omega%
} 
) \vert^2 = 
\int_0^{\Delta t}\!{\rm d} t \, \vert a( t ) \vert^2
= \alpha \frac{ \Delta x^2 }{ \Delta t^3 }
\,,
\label{eq:Parseval-Plancherel}
\end{equation}
with numerical coefficients $\alpha_n = 64, 58.51, 97.41$ for the pulse
profiles $n = 0, 1, 2$ in Fig.\:\ref{fig:force-profiles}.
This vanishes only when no splitting force is applied at all,
illustrating the fundamental character of the phonon-induced 
orthogonality in this regime.
{We finally get for the contrast}
\begin{equation}
\log C \lesssim
-
\frac{ \delta F^2 }{ F^2 }
\frac{ 3 \alpha_n }{ \pi }
\frac{ V }{ (\bar c \Delta t)^3 }
\frac{\Delta x^2 }{\lambda_1^2 }
\,.
\label{eq:many-mode-C}
\end{equation}
{We again checked the accuracy of this formula by summing over the eigenfrequencies 
of a large sphere.}
While the ratio $\delta F / F$ can be made small using high-purity material,
all other factors in Eq.\,(\ref{eq:many-mode-C}) exceed unity:
large objects are characterized by $\bar{c} \Delta t \ll L$,
and the thermal de-Broglie wavelength for a silicon atom is $\lambda_1 \approx 0.1\,{\rm nm}$ at $T = 4\,{\rm K}$, while one would aim at least for a mesoscopic splitting $
\change{
\Delta%
} 
x \gtrsim 10\,{\rm nm}$.

\begin{figure}[htbp]
\centerline{%
   \includegraphics*[width=0.9\columnwidth]{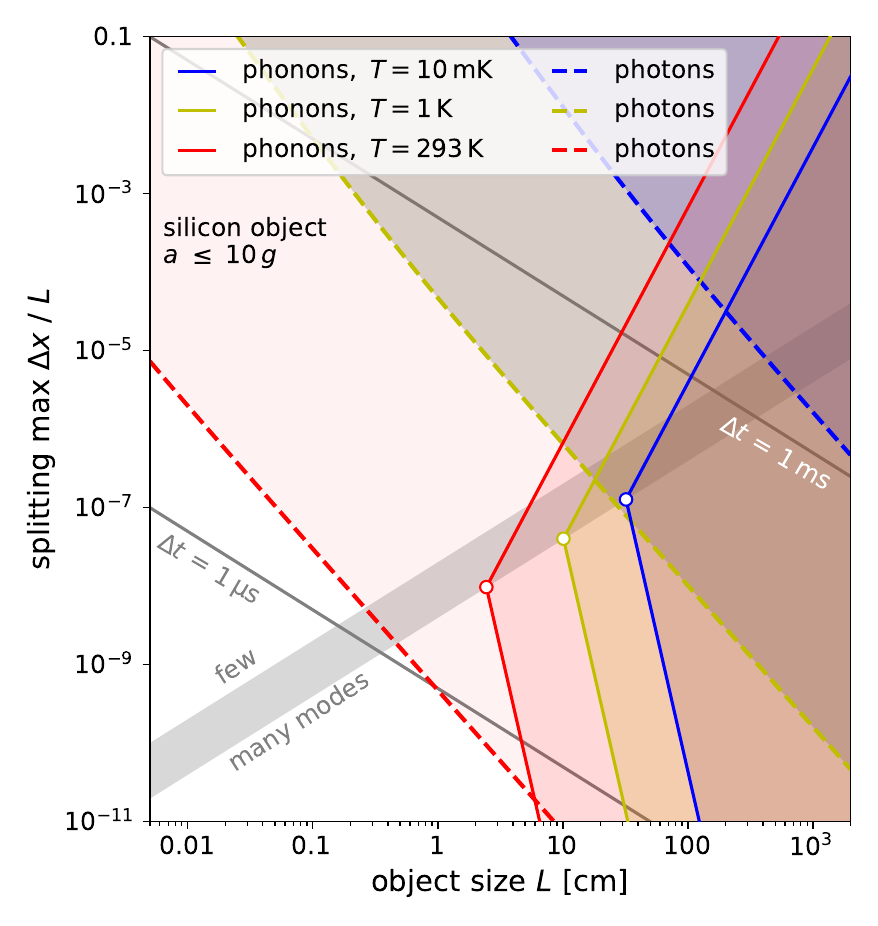}%
   }%
   \vspace*{-3ex}
   \caption[]{%
Parameter space of decoherence. Shown are lines of phonon- or photon-induced contrast reduction to $1/e$:
in the shaded areas, the contrast rapidly becomes negligible.
Thick dashed lines: decoherence due to black body radiation with temperatures $T = 293\,{\rm K}$ (red), $1\,{\rm K}$ (yellow), and $10\,{\rm mK}$ (blue), computed from the momentum diffusion constant quoted in \cite{Chang2009, Romero-Isart.2011} (see also \cite{Schlosshauer.Book} and Appendix~\ref{a:BBR}).
Solid lines with kinks:
decoherence due to phonon excitation in a spherical object at the same temperatures, assuming a force inhomogeneity of $\delta F / F = 1\%$.
The acceleration follows the protocol $a_2(t)$ of Fig.\:\ref{fig:force-profiles} and
is limited to $10\,g = 98.1\,{\rm m/s}^2$; $\Delta t$ is adjusted to achieve the given splitting (thin gray lines).
The gray band separates two
regimes where either only few phonon modes or essentially all of them
contribute (Secs.~\ref{s:few-modes}, \ref{s:many-modes}); 
it depends on the transverse speed of sound (here: $c = 5.35\,{\rm km/s}$ for silicon). More details in Appendix~\ref{a:a-omega-and-other-a's}, 
Figs.\:\ref{fig:results-2}, \ref{fig:other-accelerations}.
}
   \label{fig:results}
\end{figure}

\section{Discussion}

\subsection{Regimes of contrast loss}

In Fig.\:\ref{fig:results}, we show lines of constant contrast $C \approx 1/e$ as a function of the object size $L$ (sphere diameter) and the relative maximum splitting $\Delta x/L$.
The latter is a convenient way to parametrize the duration of the applied force (whose maximum acceleration is limited to $10\,g$ in Fig.\:\ref{fig:results}), 
while the object size $L$ gives the scaling of the phonon spectrum.
The smoothest force profile $a_2(t)$ and different object temperatures are considered here.
Further parameters are given in the caption.
The parameter sets defined by Eqs.\,(\ref{eq:log-contrast-3}, \ref{eq:many-mode-C}) derived above cross in the gray band where $\omega_1 \Delta t \sim 1$ holds.
For completeness, we provide plots for smaller maximum accelerations of $0.1\,g$ and $1\,g$ in Appendix~\ref{a:a-omega-and-other-a's}, 
Figs.~\ref{fig:results-2}, \ref{fig:other-accelerations}.
{Since splitting and recombining then take more time, isolation becomes even more of a challenge.}

Fig.\:\ref{fig:results} also shows the characteristic parameter combinations for decoherence due to black body radiation (BBR, Appendix~\ref{a:BBR}).
In the shaded areas above or to the right of the lines, the interference contrast drops below $1/e$ and rapidly becomes negligible.
When the interference experiment is performed in perfect vacuum, BBR is considered the dominant process for environmental decoherence \cite{Joos.1985,Schlosshauer.Book}.

From Fig.\:\ref{fig:results} several facts become evident.
For hot objects, BBR decoherence dominates, but this limit is not very relevant as only small spatial splittings are allowed (e.g., $10^{-11}$\,m for an object size of 1\,cm).
For cold objects, the situation is almost completely reversed.
While BBR allows for significant spatial superpositions of large objects, phonons do not.
For example, if we cool to a temperature of $1$\,K, objects the size of $1$\,m cannot be put in any spatial superposition.
The same applies to objects $10-100$\,m in size (e.g., a car or a spaceship), even if we cool to much lower temperatures.
To understand the dependence on temperature $T$, recall that the distribution function $W_k(q, \dot q)$ for amplitude quadratures of a given phonon mode $k$ is broadening as $T$ increases.
The orthogonality, however, depends on the inverse widths of this distribution, which play the role of coherence lengths 
(as in the definition of the thermal de Broglie wavelength) [see {yellow area in Fig.\:\ref{fig:orthogonality} and} Eq.\,(\ref{eq:th-eq-sigma-uv})].
At higher $T$, although many modes are already populated, phonon amplitudes with opposite signs are thus easier to distinguish, yielding more \emph{Welcher Weg} information.
Or as Lipkin phrased it: In a hotter lattice, the probability of inelastic energy exchange is increased due to stimulated emission and absorption of phonons
\cite{Lipkin.1960b}.
Finally, Fig.\:\ref{fig:results-2} in the Appendix shows that the more adiabatic the force profiles are, the weaker the phonon-induced decoherence is.

\subsection{Universal and fundamental features}

We believe the orthogonality formation described in this work is universal: why? 
Coherent spatial splitting can either be achieved by active splitting (e.g., Stern-Gerlach force or light force) or be the result of a passive procedure
(e.g., a coherence slit or a beam expander before the two slits in a double-slit experiment).
The relevance of our calculation to any active splitting is clear, so
let us now discuss passive splitting.
Consider the two wave packets that reach the left or right slit after a coherence slit.
If we believe in momentum conservation, the coherence slit must have received some recoil momentum along its plane.
Quantum mechanics describes this as entanglement between the momenta of the coherence slit and of the particle.
It is legitimate to consider that the momentum transfer originates from some force.
In Bohmian mechanics, it would be due to the quantum potential, but this is not important.
We may then ask if this force acts identically on all atoms within the particle going through the slit.
This can be tested experimentally with a cold object, phonons frozen out, diffracted by a single slit.
The width of the slit must be larger than the object, and the object's transverse coherence length should be about the size of the slit.
We may then capture the particle after it has been scattering by some non-negligible angle, and check whether phonons have been excited.
If this is the case, then it stands to reason that scattering to the left or to the right correlates with a different excitation of phonons (at least in phase, as symmetry was broken).
In that scenario, the calculations of the paper are also valid for passive splitting, and the effect may be termed universal.

Finally, let us argue that the effect is fundamental, or at the very least as fundamental as the arrow of time.
It is clear that in principle all phase-space dynamics may be time-reversed.
However, for all practical purposes this becomes harder and harder as the Hilbert space increases in size, or as the number of DoF increases.
Specifically to our case, we believe it is impractical to try and reverse the evolution of the excited phonons in a massive particle, due to the lacking possibility of addressing modes efficiently and due to their sheer number.
For example, if we take a 0.5\,m object so that the fundamental wavelength is 1\,m, and if we take a speed of sound of 1000\,m/s we find a kHz frequency.
This means that if we give a 1\,ns force pulse, it will excite $10^6$ phonons or $10^{18}$ in 3D.
These are huge numbers to follow and manipulate.
Typically, the shape of the particle will be such that these phonons will not have some magical frequency ratio which would enable some natural rephasing.
Obviously, non-linearities such as phonon-phonon scattering will complicate things even further.
We therefore believe that even in an isolated system, if it is large enough,
the internal DoF engage in the same scenario as an external environment and make the time evolution, for all practical purposes, irreversible, as it happens with the thermodynamic arrow of time.

\section{Conclusion}
\label{s:conclusion}

In this paper, we have argued that vibrations internal to any object, mesoscopic or larger, provide a way to {leak} \emph{Welcher Weg} information in an interferometer based on spatially split arms.
This information makes the wave packets along two interferometer arms orthogonal in the multi-mode phase space of the center of mass dynamics compounded with internal vibrations.
This cannot be suppressed by isolation from the external environment and puts stringent constraints on future spatial superpositions of large objects.
Significant contrast reduction may happen, as soon as a single mode is excited by the splitting force into orthogonal states in its phase space. 
But it also appears when many modes, typical for large objects, suffer only a small differential displacement. 
We evaluated the corresponding product over internal modes and found that for sufficiently large objects (see Fig.\:\ref{fig:results}), 
\emph{Welcher Weg} information encoded in phonons is in practice impossible to erase. 
The scaling with object size and its internal temperature has been computed and provides, for certain object sizes and splittings, bounds for coherent interferometry that are more stringent than those due to black body photons.

Methods may be found in the future
to overcome this limit.
Let us give a few examples: First, the applied force may be made more homogeneous. This requires not only
a highly constant potential gradient,
but also a very pure and homogeneous material.
An example for such a force is gravity, as long as the particle size is small enough to neglect tidal forces.
Next, a high level of adiabaticity may be achieved, although this typically also requires long durations, meaning that other decoherence channels, e.g., black body radiation, will become strong. 
Alternatively, one may engineer
pulse spectra whose zeros [seen in Fig.\:\ref{fig:force-profiles}(B)]
overlap with the relevant phonon frequencies,
assuming that there are just a few of them (small-object regime).
An additional path may include a designed geometrical structure for the object which prohibits low-lying phonons, in other words, a phononic bandgap material. 
Last, one may hope to reverse the evolution of the phonons, which in principle is allowed by time-reversibility, but may require sophisticated control tools to address them at the single-phonon quantum level.

Finally, we believe the limit presented here to be {\it universal}, namely that it applies to all splitting procedures whether active (i.e., a force is applied, as calculated above), or passive (i.e., as in a single slit creating diffraction and momentum separation before a double-slit screen). 
We also argued that the effect may be termed {\it fundamental}, or at least, as fundamental as the arrow of time, whereby time-irreversibility emerges for large systems. This will preclude, for all practical purposes, the reversal of the phonon evolution in a large system.

\bigskip

\begin{acknowledgments}

This work is funded in part by the Israel Science Foundation Grant Nos.\ 856/18,
1314/19, 3515/20, 3470/21. This work was partly supported by the Gordon and Betty Moore Foundation 
(doi.org/10.37807/GBMF11936), and the Simons Foundation, via the Small-Scale Experiments for Fundamental Physics program.
This research was also partly supported by the Israel Innovation Authority
(Grant No.\ 76974)
as part of the Horizon Europe QuantERA program (LEMAQUME project).
This research was supported in part by the National Science Foundation (NSF) under Grant No.\ PHY-1748958
and fueled by the Deutsche Forschungsgemeinschaft (DFG, German Research Foundation) within SFB/CRC 1636, ID 510943930, Project No.\ A01.
C.H.\ thanks the KITP (University of California at Santa Barbara) for hospitality during the program ``Quantum and Thermal Electrodynamic Fluctuations in the Presence of Matter: Progress and Challenges''.
He acknowledges discussions with Ph.\ Richter and L. Saviot.
\end{acknowledgments}

\appendix

\section{Decoherence parameter space for other accelerations}
\label{a:a-omega-and-other-a's}

The three generic protocols for the applied splitting force 
that are sketched in Fig.\:\ref{fig:force-profiles} 
have Fourier transforms defined in Eq.\,(\ref{eq:scaling-a-omega})
%
\begin{align}
\accVarphi_0(\omega \Delta t) &=
\frac{ 16 }{ \omega \Delta t }
\left[ \sin (\omega  \Delta t / 2) - 2 \sin(\omega \Delta t / 4) \right]
\,,
\label{eq:a0-spectrum}
\\
\accVarphi_1(\omega \Delta t) &=
- \frac{ 96 }{ (\omega \Delta t/2)^2 } \left[
\cos(\omega \Delta t / 2)
\left( 1 - \frac{60}{(\omega \Delta t)^2} \right)
\right.
\nonumber\\
& \qquad
\left. {}
-
\frac{ \sin(\omega \Delta t / 2) }{ \omega \Delta t / 2 }
\left( 6 - \frac{60}{(\omega \Delta t)^2} \right)
\right]
\,,
\\
\accVarphi_2(\omega \Delta t) &=
- 3 \pi^4 \frac{\sin(\omega \Delta t / 2)}{(\omega \Delta t / 2)^3}
\label{eq:spectrum-acceleration}
\\
& \qquad
\times
\left( 1 - \frac{ (2 \pi)^2 }{ (\omega \Delta t)^2 }\right)^{-1}
\left( 1 - \frac{ (4 \pi)^2 }{ (\omega \Delta t)^2 }\right)^{-1}
\nonumber
\,.
\end{align}
They produce the same spatial splitting at mid-time $\Delta t/2$,
and have zero time-average (in order to close the loop for the CoM), 
but differ in the degree of discontinuity as the force is switched on and off. 
The profile $\accVarphi_0$, e.g., is rectangular and least adiabatic.

\begin{figure}[tbp]
\centerline{%
   \includegraphics*[width=0.9\columnwidth]{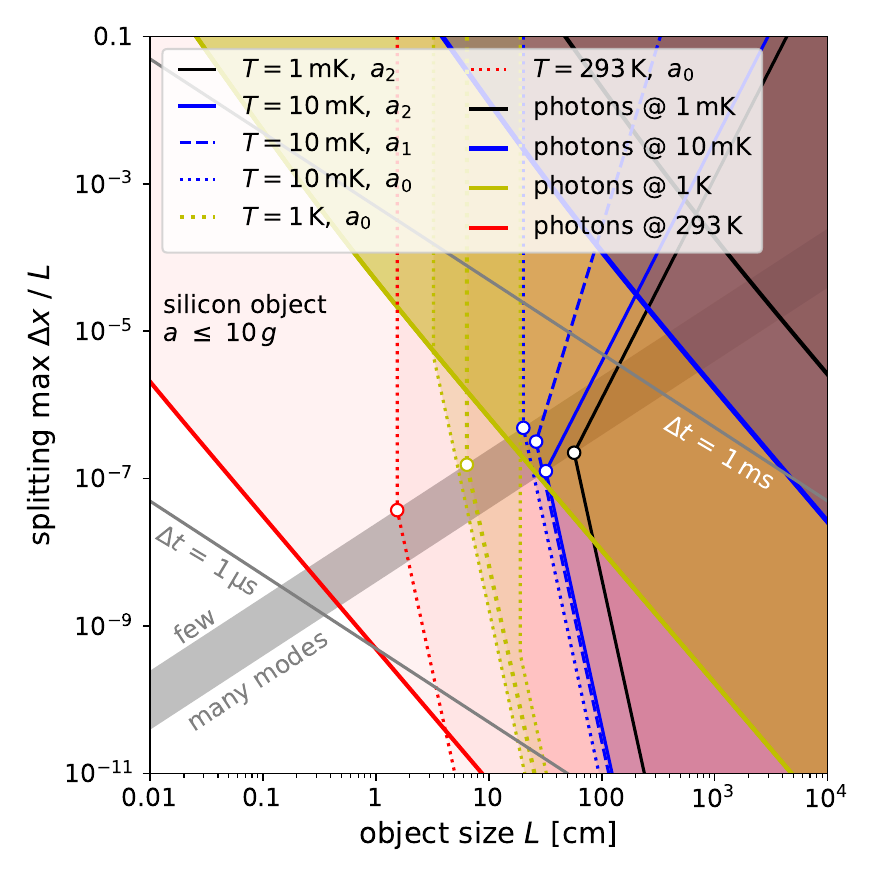}%
   }%
   \vspace*{-3ex}
   \caption[]{%
Further details on the parameter space of decoherence, see Fig.\:\ref{fig:results}. Shown are lines of phonon- or photon-induced contrast reduction to $1/e$:
in the shaded areas, the contrast rapidly becomes negligible.
Thick colored lines: decoherence due to black body radiation with temperatures $T = 293\,{\rm K}$ (red), $1\,{\rm K}$ (yellow), $10\,{\rm mK}$ (blue), and $1\,{\rm mK}$ (black).
Dotted (dashed, solid) lines with kinks:
decoherence due to phonon excitation in a spherical object at the same temperatures, assuming a force inhomogeneity of $\delta F / F = 1\%$.
The line style corresponds to the applied acceleration protocols
$a_{0}(t) \ldots a_2(t)$ shown in Fig.\:\ref{fig:force-profiles}.
The gray band separates the regimes
where few phonon modes or essentially all of them
contribute (Sec.\,3).
The family of dotted yellow lines ($T = 1\,{\rm K}$) represents the effect of the (transverse) speed of sound: steel (left edge: $3.2\,{\rm km/s}$), silicon (center: $5.35\,{\rm km/s}$), and diamond (right edge: $12.8\,{\rm km/s}$, Ref.\,\cite{Wang.2004}).
}
   \label{fig:results-2}
\end{figure}

The degree of adiabaticity has strong implications for the contrast reduction, as can be seen in Fig.\:\ref{fig:results-2}.
Scaling laws with different exponents emerge in the few-mode limit, due to the way high frequencies are suppressed in the spectra $\accVarphi_n(\omega \Delta)$.
Indeed, one can read off from Eqs.\,(\ref{eq:a0-spectrum}--\ref{eq:spectrum-acceleration}) that the Fourier transforms $\accVarphi_n(\omega \Delta t)$ follow approximately power laws with exponents $-(n+1)$, $n = 0, 1, 2$, 
as one considers large enough frequencies, i.e., $\omega \Delta t \gg 1$.
In the limit that many modes contribute, a more adiabatic pulse provides less
advantages (see Fig.\:\ref{fig:results-2}).
This can be attributed to the ``sum rule'' of Eq.\,(\ref{eq:Parseval-Plancherel}) 
where the three profiles only differ in their overall amplitude $\alpha_n$.

\begin{figure*}[htbp]
   \centering
   \includegraphics[height=0.45\textwidth]{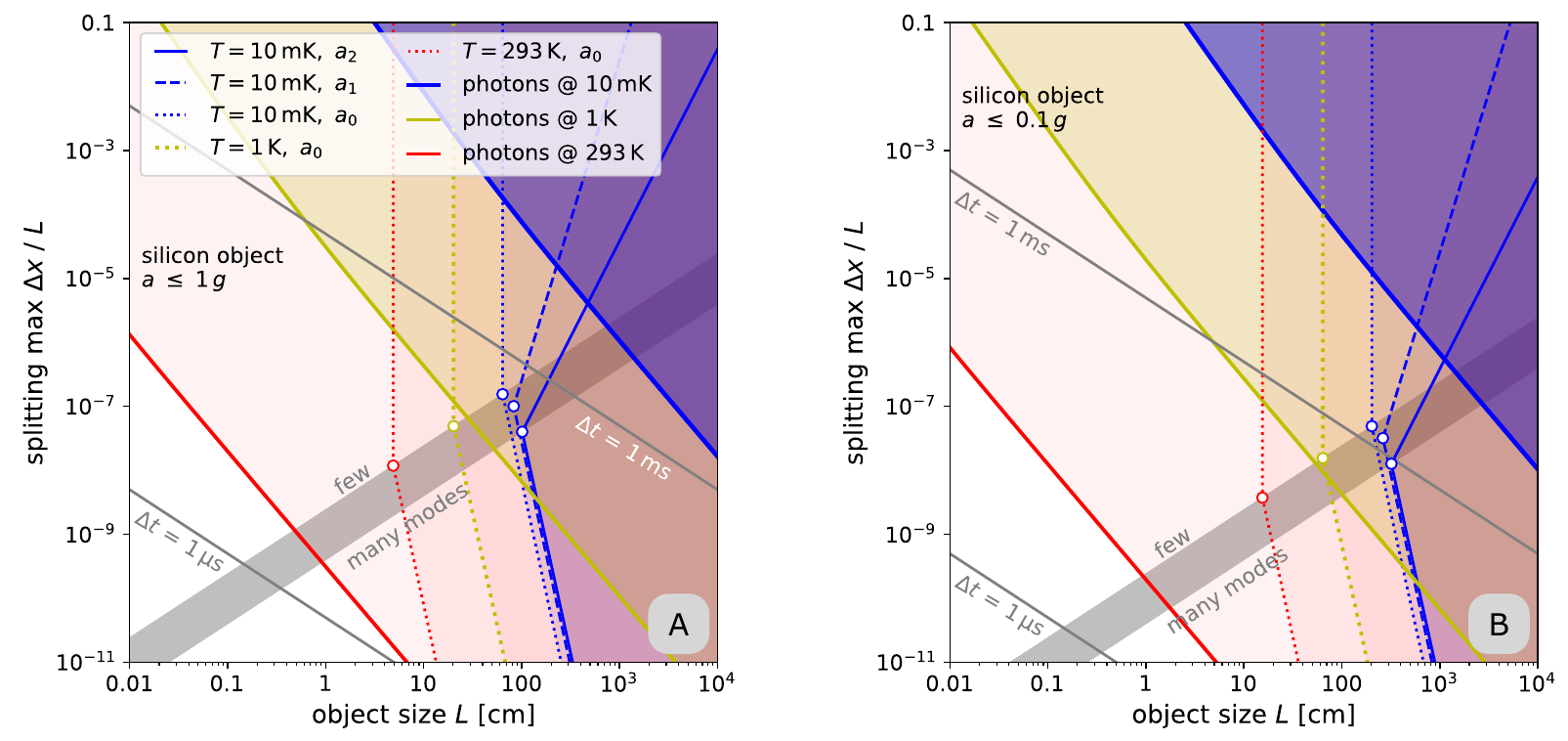}
   \caption[]{Same as Fig.\:\ref{fig:results-2}, but for smaller maximum accelerations, limited to $1\,g$ (A) and $0.1\,g$ (B) with terrestrial gravity $g = 9.81\,{\rm m/s}^2$. The data with $T = 1\,{\rm mK}$ and other materials are left out for clarity. To reach the targeted maximum splitting at smaller accelerations, a longer force sequence is needed (slanted grey lines).}
   \label{fig:other-accelerations}
\end{figure*}

Figure\,\ref{fig:results-2} illustrates the impact of choosing less adiabatic force pulses on the loss of interference contrast,
to be compared to Fig.\:\ref{fig:results}.
The dotted yellow curves illustrate results for three materials 
(with different speed of sound). 
Finally, 
Fig.\:\ref{fig:other-accelerations} explores the critical parameter space 
with smaller maximum accelerations.

\section{Decoherence from black body photons}
\label{a:BBR}

Here we discuss decoherence due to black body radiation (BBR).
We take into account the momentum exchange due to absorption and emission of photons and characterize it by the diffusion coefficient $D_p$ of Schlosshauer \cite{Schlosshauer.Book} and Romero-Isart \cite{Romero-Isart.2011}.
The limiting cases of a small Rayleigh scatterer (radius smaller than the Wien wavelength) and of a large object (absorption given by the geometrical cross section) are interpolated
with a Pad\'e approximation.
The optical constants for the infrared absorption are taken from Ref.\,\cite{Romero-Isart.2011}.
The contrast reduction for the case of a time-dependent applied force is computed from the integral
\begin{equation}
\log C_{\rm BBR} = - \int_0^{\Delta t}\!{\rm d}t\, \frac{ D_p }{ \hbar^2 }
\Delta x^2(t)
\,,
\label{eq:spatial-decoherence}
\end{equation}
\noindent where $\Delta x(t)$ is the time-dependent spatial separation between the two wave packets.
Equation~(\ref{eq:spatial-decoherence}) provides the washing-out of the interference fringes (“ghost component”) of the Wigner function for a superposition of wave packets \cite{Schleich.Book, VogelWelsch.Book}, due to momentum diffusion as thermal photons are absorbed or emitted.
This is valid as long as $\Delta x$ remains small compared to the coherence length of the black body field, itself of the order of the Wien wavelength~\cite{Cheng.1999, Romero-Isart.2011}.
The integral can be evaluated analytically for the applied force protocols of Fig.\:\ref{fig:force-profiles}.



%

%

\end{document}